\documentclass[prl,twocolumn,showpacs,amsmath,superscriptaddress]{revtex4}

\usepackage{graphicx}
\usepackage{amssymb,amsmath}
\usepackage{dcolumn}
\usepackage{bm}
\usepackage{color}

\begin{document}

\title{Bose glass and superfuid phases of cavity polaritons}
\author{G. Malpuech}
\affiliation{LASMEA, CNRS-Universit\'e Blaise Pascal, 24 Avenue des Landais, 63177 Aubi\`ere Cedex France}
\author{D. D. Solnyshkov}
\affiliation{LASMEA, CNRS-Universit\'e Blaise Pascal, 24 Avenue des Landais, 63177 Aubi\`ere Cedex France}
\author{H. Ouerdane}
\affiliation{LASMEA, CNRS-Universit\'e Blaise Pascal, 24 Avenue des Landais, 63177 Aubi\`ere Cedex France}
\author{M. M. Glazov}
\affiliation{A. F. Ioffe Physico-Technical Institute, Russian Academy of Sciences, 194021 St. Petersburg, Russia}
\author{I. Shelykh}
\affiliation{ICCMP, Universidade de Bras\`{\i}lia, 70919-970 Bras\`{\i}lia-DF, Brazil}
\date{\today}

\begin{abstract}
We report the calculation of cavity exciton-polariton phase diagram which
takes into account the presence of realistic structural disorder. Polaritons
are modelled as weakly interacting two-dimensional bosons. We show that with
increasing density polaritons first undergo a quasi-phase transition toward
a Bose glass: the condensate is localized in at least one minimum of
the disorder potential, depending on the value of the chemical potential of
the polariton system. A further increase of the density leads to a
percolation process of the polariton fluid which gives rise to a
Kosterlitz-Thouless phase transition towards superfluidity. The spatial
representation of the condensate wavefunction as well as the spectrum of
elementary excitations are obtained from the solution of the
Gross-Pitaevskii equation for all the phases.
\end{abstract}

\maketitle


Exciton-polaritons (polaritons) in microcavities are composite 2-dimensional
weakly interacting bosons \cite{ref1, Imamoglu}. Despite their short radiative life
time ($\sim 10^{-12}$ s) stimulated scattering towards their
ground state has been demonstrated \cite{Dang98,Yamamoto}, and
quasi-thermal equilibrium was recently reported \cite{KAZ06}.
In this quasi-equilibrium regime cavity
polaritons are expected to give rise to a Kosterlitz-Thouless (KT) phase
transition toward superfluidity \cite{KOS73}. The corresponding phase
diagram has been established a few years ago \cite{MAL03}, and recently
refined to fully take into account the non-parabolic shape of the polariton
dispersion \cite{KEE06}. Because of their light effective mass (typically $%
10^{-4}$ times the free electron mass) polaritons show extremely small
critical density and high critical temperatures that can be larger than room
temperature in some cases. However, semiconductors were assumed to be ideal
in the approach used in Refs.~\cite{MAL03} and \cite{KEE06} while
experimental data clearly show strong localization of the condensate because
of structural imperfections. The phase observed is in fact characteristic of
a Bose glass \cite{FIS89} and no signature of superfluidity has been
reported thus far. In this Letter we propose the derivation of a new
polariton phase diagram taking into account structural disorder whose impact
on the spatial shape of the wavefunction and the dispersion of elementary
excitations, is analyzed within the framework of the Gross-Pitaevskii theory
\cite{Pitaevskiibook}.

To give a qualitative picture of the model, we assume that the polaritons
are moving in a random potential $V(\mathbf{r})$ whose mean amplitude and
root mean square fluctuation are given by $\langle V(\mathbf{r})\rangle =0$
and $\sqrt{\langle V^{2}(\mathbf{r})\rangle }=V_{0}$ respectively. The
correlation length of this potential is $R_{0}=\sqrt{\int \langle V(%
\mathbf{r})V(0)\rangle \mathrm{d}\mathbf{r}/V_{0}^{2}}$. As in any 
disordered system, there are here two types of polaritons states ~\cite{EFR89}: 
the free propagating states and the localized states with energy $E<E_{\mathrm{c}}$, 
where $E_{\mathrm{c}}$ is the critical \textquotedblleft delocalization\textquotedblright\ 
energy. The localization radius scales like $a(E)\propto a_{0}V_{0}^{s}/(E_{\mathrm{c}%
}-E)^{s}$, $s$ being a critical index and $a_{0}=\sqrt{\hbar
^{2}/mV_{0}}$ \cite{explain1}. In two dimensions $E_{\mathrm{c}}$ is of 
the order of mean potential energy (i.e. $0$ in our
case), and $s\approx 0.75$~\cite{EFR89}. The quasi-classical density of
states is $D_{0}(E)\cong M/4\pi \hbar ^{2}[1+\mathop{\rm erf}\nolimits{(E/V_{0})}]$
\cite{explain2}.
However, an exciton-polariton 
is a composite boson also containing a photonic component which makes
its trapping in the state with the localization radius
$a=a_{\mathrm{c}}\lesssim \lambda $ (where $\lambda $ is the
wavelength of the incident light) not possible. Thus the
resulting effective density of states shows an abrupt
cut-off at $E=E_{0}$ which is self-consistently determined from $a(E_{0})=a_{%
\mathrm{c}}$: $D(E)=D_{0}(E)$ for $E>E_{0}$ and $D(E)=0$
otherwise.

Clearly, non-interacting bosons cannot undergo Bose Einstein Condensation (BEC)
as the number of particles which can be fitted to all
the excited states of the system ($E>E_0$) is divergent.
The situation is thus different from
those for cold atoms trapped by power-low potential in 2D, where
the renormalization of the density of states makes ``true'' BEC
possible \cite{Bagnato}. Therefore, even in the presence of
disorder, BEC cannot take place strictly speaking
for cavity polaritons. However, it is possible to define a
quasi-phase transition which takes place in finite systems
\cite{MAL03}. Indeed, for the finite size $R$ system there is a
finite number $N_{\mathrm{trap}}$ of potential traps for
polaritons, thus, there is an energy spacing between the single particle
states. The typical energy distance between the ground and excited
states of the finite-size system levels $\delta $ under the
assumption of long-range potential is approximately given by
$V_{0}/N_{\mathrm{trap}}$ or $\hbar ^{2}/2MR_{0}^{2}$ whichever is
smaller. In this framework the critical density is given by the total number of
polaritons which can be accomodated in all the energy levels of the disorder
potential $V(\mathbf{r})$ except the ground one \cite{MAL03}:

\begin{equation}
n_{\mathrm{c}}(T)=\sum_{i\neq 0}f_{\mathrm{B}}(E_{i},E_{0},T),
\label{critical}
\end{equation}
where $f_{\mathrm{B}}(T,\mu ,T)$ is the Bose-Einstein distribution
function, $\mu $ is the chemical potential.

To evaluate the critical density $n_{\mathrm{c}}(T,R)$, the discrete sum is
replaced by an integral in Eq.~\eqref{critical}, and we find $n_{\mathrm{c}%
}(T)\approx D(E_{0})k_{\mathrm{B}}T\ln \left[1/(1-e^{\delta /k_{\mathrm{B}}T})\right]$ 
assuming $D(E)$ is a smooth function. Above this density all additional particles are
accumulating in the ground state and the concentration of condensed
particles $n_{0}$ satisfies $n_{0}\geq (n-n_{\mathrm{c}})$, where $n$ is the
total density of polaritons. It is not a real phase transition since the
system has a discrete energy spectrum and the value of the chemical
potential never becomes strictly equal to $E_{0}$.

Interactions between particles start to become dominant
once the polaritons start to accumulate in the ground state. The
situation can be qualitatively described as follows: particles
start to fill the lowest energy state which is therefore blue
shifted because of interactions ($\mu -E_{0}>0$). Thus, for 
some occupation number of the condensate the chemical
potential reaches the energy of another localized state, and this
state starts in turn to populate and to blue shift.
The condensate, like a liquid, fills several minima
of the potential. It gives rise to the spatial and reciprocal
space pictures of Refs.~\cite{KAZ06,KOS73}. A few localized
states, covering about 20 \% of the surface of the emitting spot
are all emitting light at the same energy and are
strongly populated. This characterizes a Bose glass \cite{FIS89}.
This situation occurs up to the achievement of the condition $\mu
=E_{\rm c}$. This condition should be accompanied by a percolation of
the condensate which at this stage should cover 50 \%
or more of the sample (in the semiclassical representation). The delocalized
condensate becomes at this stage a KT superfluid. More precisely,
the different sides of the finite-size system are linked by the
phase coherent path. Therefore we predict \emph{two} quasi-phase
transitions driven by temperature and particle density: 
first, with an increase of the polariton density beyond 
$n_{\rm c}(T)$ the system enters the Bose glass phase, then with a
further increase of the density the polariton system becomes
superfluid. The critical condition $\mu =E_{\rm c}$ is valid only at
low temperature where the thermal depletion of the condensate is
negligible.

The quantitative analysis can be carried out in the framework of the
Gross-Pitaevskii equation for the condensate wavefunction
$\Psi (\mathbf{r},t)$ which reads

\begin{equation}
i\hbar \frac{\displaystyle\partial }{\displaystyle\partial t}\Psi (\mathbf{r}%
,t)=\left( -\frac{\displaystyle\hbar ^{2}}{\displaystyle2M}\triangle +V(%
\mathbf{r})+g\left\vert \Psi (\mathbf{r},t)\right\vert ^{2}\right) \Psi (%
\mathbf{r},t),  \label{gp1}
\end{equation}%

\noindent where $g$ is a constant characterizing the weak repulsive interaction
between polaritons. The solution of the Gross-Pitaevskii equation takes the form $\Psi (\mathbf{r%
},t)=\Psi _{0}(\mathbf{r})\exp \left( -i\mu t/\hbar \right) $. Top panels (a), (b), and (c) of Fig.~1 show
the real space distribution of the polaritons obtained from the
solution of the Gross-Pitaevskii equation. The parameters are those of a realistic CdTe
microcavity at zero detuning. We have taken the polariton mass 
$m=5 \times 10^{-5}m_{0}$, where $m_{0}$ is the free electron mass, 
and the interaction constant $g=3E_{\rm b}a_{\rm B}^{2}/N_{\rm {qw}}$,
where $E_{\rm b}$ is the exciton binding energy (25 meV in CdTe), $a_{\rm B}=34~ 
\mathring{A}$ is the exciton Bohr radius and $N_{\rm qw}=16$ is the number of wells
embedded in the microcavity. We have included a
random disorder potential with $V_{0}=0.5$ meV
and $R_{0}=3$ $\mu $m. Figure 1.a corresponds
to the non-condensed situation. The spatial profile is
given by the statistical averaging over the all occupied states, $n(\mathbf{r%
})=\sum_{j}f_{B}(E_{j},T,\mu (T))|\Psi _{j}(\mathbf{r})|^{2}$. Here the
temperature is set to $T=19$ K which corresponds to the effective polariton temperature 
measured in \cite{KAZ06}. In this case the total number of particles is small
and thus nonlinear terms in Gross-Pitaevskii equation can be neglected. 

Once the quasi-condensate is formed, and for moderate temperatures, one can
neglect the thermal occupation of the excited states and the spatial image
of the polariton distribution is given by the ground state wavefunction
which corresponds to solution of Eq.~\eqref{gp1}. We show the resulting density 
below and above the percolation threshold on Figs. 1.b and 1.c respectively.
As expected, the condensate is localized in a few minima of the random potential as shown on Fig.~1.b.
On Fig.~1.c the condensate wave function still exhibits some spatial fluctuations connected to
disorder, but the condensate is nonetheless well delocalized, covering the whole sample area.

\begin{figure}
\includegraphics[width=0.95\linewidth]{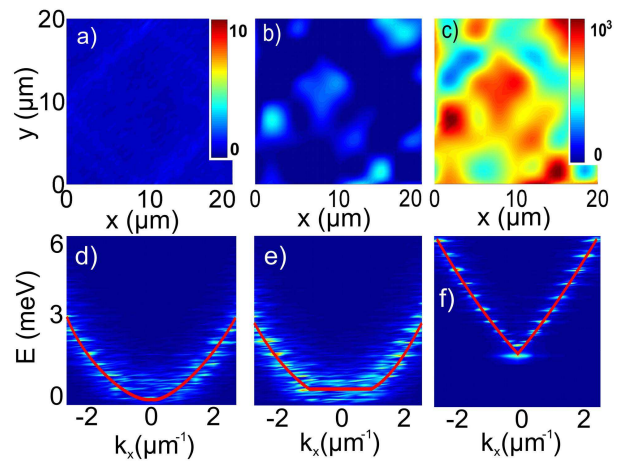}
\caption{(Color online) Spatial images (top panels) and quasiparticle spectra
(bottom panels) for a realistic disorder potential. The figures
shown correspond to densities 0, $6\times 10^{10}$ and $2\times
10^{12}$ cm$^{-2}$. The red lines are only guides to the eye, 
showing parabolic, flat and linear-type dispersions. 
The colormap of the panel 1.b is the same as 1.c.} \label{fig1}
\end{figure}

To calculate the quasiparticle spectra shown in
lower panels 1.d, 1.e and 1.f of Fig. 1 we introduce a
single-particle Green's function which takes the form

\begin{equation}
G_{\omega }(\mathbf{r},\mathbf{r}_{0})=\sum_{j}\frac{\Psi _{j}(\mathbf{r}%
)\Psi _{j}^{\dag }(\mathbf{r}_{0})}{\hbar\omega -E_{j}},  \label{greens}
\end{equation}

\noindent where \emph{E}$_{j}$ and  $\Psi _{j}(\mathbf{r})$ are energies and
eigenfunctions of the elementary excitations \cite{pitaevskii}, 
found numerically from Eq.~\eqref{gp1}. The spectrum
of elementary excitations is given by the poles of the Green's
function in the (\textbf{k}, $\omega$) representation, and shown on 
the lower panels of Fig.~1. The left panel 1.d shows typical parabolic
dispersion broadened by the disorder potential. The middle panel
1.e shows parabolic dispersion with a flat part produced by the
localization of the condensate. The linear spectrum on the right
panel 1.f is the distinct feature of the superfluid state of the
system. Only the upper Bogoliubov branch is shown. Figures 1.b and 1.e
reproduce quite well the experimental observations of Ref.~\cite{KAZ06} which
are characteristics of the formation of a Bose glass.

It is instructive to analyze both the variation of the emission pattern
and the quasi-particle spectrum in comparison with the behavior of the superfluid 
fraction of the polariton system. The latter quantity can be calculated using the twisted
boundary conditions method~\cite{leggett}. Imposing such boundary conditions 
implies that the condensate wavefunction acquires a phase between the boundaries, namely

\begin{equation}  \label{twisted}
\Psi_\theta(\mathbf{r}+ \mathbf{L}_i) = \mathrm{e}^{\mathrm{i
}\theta} \Psi_\theta(\mathbf{r}),
\end{equation}

\noindent where $\mathbf{L}_i$ ($i=x,y$) are the vectors which form the
rectangle confining the polaritons and $\theta$ is the twisting
parameter. The superfluid fraction of the condensate is given by~\cite{leggett}

\begin{equation}  \label{fs}
f_{\rm s} = \frac{n_{\rm s}}{n} =
\lim_{\theta\to 0} \frac{2ML^2(\mu_\theta - \mu_0)}{%
\hbar^2 n \theta^2},
\end{equation}

\noindent where $\mu_{\theta}$ is the chemical potential corresponding to
the boundary conditions Eq. \eqref{twisted} and $\mu_0$ is the
chemical potential corresponding to the periodic boundary
conditions ($\theta=0$). In the case of a clean system, $V(\mathbf{r}) = 0$, 
the plane wave is the solution of Eq.~\eqref{gp1} and 
$\mu_{\theta} - \mu_0 = n\hbar^2\theta^2/2ML^2$:
the superfluid fraction is $f_{\rm s}= 1$. On the contrary,
for the strongly localized condensate the wavefunction is
exponentially small at the system
boundaries and the change of the boundary condition (i.e. variation of $%
\theta$) does not change the energy of the system, thus $f_{\rm s} \sim \exp{%
[-L/a(\mu_0)]}$ and goes to $0$ for the infinite system. Due to
the exponential tails of the localized wavefunctions a small
degree of superfluidity remains in the finite size system. Equations 
\eqref{gp1} and \eqref{twisted} allow to study the depletion of
the superfluid fraction for arbitrary disorder. The
contribution of the disorder to the normal density of polaritons
can be represented as

\begin{equation}  \label{nnd}
n_{\mathrm{n}}^{\mathrm{d}} = (1-f_{\rm s}) n.
\end{equation}

\begin {figure}[!rh]
\centering
\scalebox{.67}{\rotatebox{0}{\includegraphics*{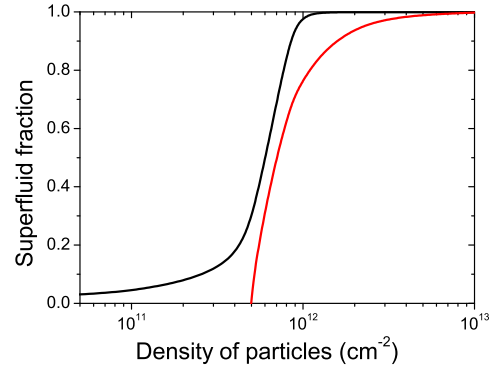}}}
\caption{(Color online) Superfluid fraction as a function of the density of
particles, obtained from twisted boundary conditions (black curve)
and from the perturbative approach (red curve).} \label{fig2}
\end{figure}

Figure~2 shows the superfluid fraction calculated as a function of
the polariton density in the system for $T=0~\mbox{K}$. Due to
the finiteness of the system considered the superfluid fraction
remains non-zero for any finite density, but a very
clear threshold behavior for densities corresponding to the
percolation threshold as observed on Fig.~1, is also shown.
For high values of the chemical potential, where $V_0^2/\mu g \ll 1$,
perturbation theory applies and we obtain

\begin{equation}\label{8}
n_{\mathrm{n}}^{\mathrm{d}} =  \frac{1}{4 } \frac{V_0^2}{\mu g},
\end{equation}

\noindent for the normal density~\cite{BER04}. The resulting curve 
is shown in red on Fig.~2. The twisted boundary
conditions approach and Eq. \eqref{8} give coinciding results for
high density of the polaritons.

In the rest of the paper, we concentrate on the establishment of
the cavity polariton phase diagram. Similarly to previous
works \cite{MAL03}, we roughly define a temperature and density
domain where the strong coupling is supposed to hold. The limits are
shown on Fig.~3 as thick dotted lines \footnote{The edge temperature is 
assumed to be 300 K equal to the exciton
binding energy. The maximum polariton density is taken 
32 times larger than bleaching exciton
density which is assumed to be $10^{11}$ cm$^{-2}$ in CdTe.}. The
transition from normal to Bose glass phase can be calculated
from Eq.~\eqref{critical} and a realistic realization of disorder. 
The lower solid line on Fig. 3 shows $n_{\mathrm{c}}(T)$ for the 
same realization of disorder as for Fig.~1.
The free polariton dispersion is calculated using the geometry of Ref.~\cite{KAZ06}.
At $T = 19$ K we find $n_{\mathrm{c}} =$ $2\times 10^8 \mbox{cm}^{-2}$.
 
We now calculate the density for the transition between the
Bose glass and the superfluid phase. In the low temperature domain,
this density is approximately given by the percolation threshold $\mu
=E_{c}$ and does not depend significantly on temperature. This
condition corresponds with good accuracy to the abrupt change of
the superfluid fraction $f_s$ shown in Fig.~2. However, at higher temperature the
thermal depletion of the condensate becomes the dominant effect.
In that case the chemical potential of the condensate is much
higher than the percolation energy $E_{\rm c}$ and the depletion induced by disorder
can be neglected compared to the thermal depletion of the superfluid. 
The normal density then reads
 
\begin{equation}  \label{eq6a}
n_{\mathrm{n}}^0(T) = - \frac{\displaystyle 2}{\displaystyle (2\pi)^2}~\int E(%
\mathbf{k})~\frac{\displaystyle \partial f_{\mathrm{B}} \left[\epsilon(%
\mathbf{k}),\mu=0,T\right]}{\displaystyle \partial \epsilon}~\mathrm{d}%
\mathbf{k},
\end{equation}

\noindent and the superfluid density in the system is given by

\begin{equation}\label{nsfinal}
n_{\mathrm{s}}(T) = n - n_{\mathrm{n}}^0(T),
\end{equation}

\noindent which can be substituted into the Kosterlitz-Nelson formula~\cite{KN}
to obtain a self-consistent equation for the transition temperature:

\begin{equation}  \label{eq8}
T_{\mathrm{KT}} = \frac{\hbar^2 \displaystyle \pi n_{\mathrm{s}}(T_{\mathrm{KT}})}{%
\displaystyle 2M}.
\end{equation}

\noindent The joint solution of Eqs. \eqref{eq8} and Eq. \eqref{nsfinal}
allows to determine the superfluid phase transition temperature
$T_{KT}(n)$. The result of this procedure is shown on Fig.~3.
Below 120 K the critical density is given by the percolation
threshold and there is no temperature dependence. Above 200 K the
superfluid depletion is determined solely by the thermal effects.
In the intermediate regime the crossover between the thermal and
disorder contributions takes place and our approximations are no longer justified.
We also find that the superfluid transition takes place very close to
the weak to strong coupling threshold and for densities 3 orders
of magnitude larger than the one of the Bose glass transition at
19 K. This suggests that experimental observation of this 
phenomenon remains a great challenge. 

\begin {figure}[!rh]
\centering
\scalebox{.57}{\rotatebox{0}{\includegraphics*{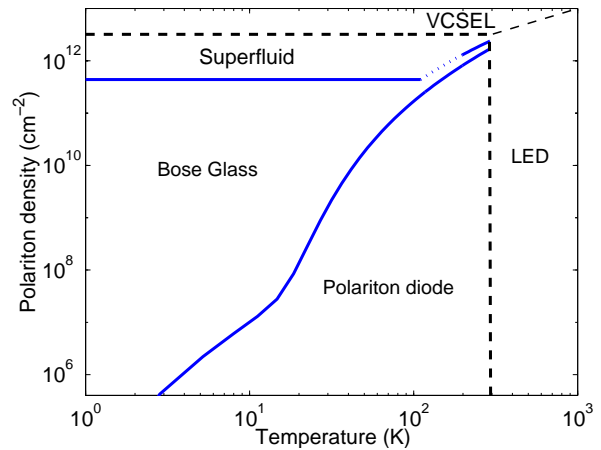}}}
\caption{(Color online) Polariton phase diagram for a CdTe microcavity containing 16 QWs.
The horizontal and vertical dashed lines show the limiting temperatures and
densities where the strong coupling holds. The lower solid line show the critical
density for the transition from normal to Bose glass phase. The upper
solid line shows the critical density for the transition from the Bose glass
to the superfluid phase. The dashed part of the line shows the temperature range  
where the validity of our approximations ceases.}
\label{fig3}
\end{figure}

In conclusion, we have established the phase diagram of cavity polaritons
taking into account the effect of structural imperfections. We predict that with
increasing density the polariton system first enters the Bose glass phase before
it becomes superfluid. The Bose glass picture is in good agreement with recent 
experimental data \cite{KAZ06}. The condensate wavefunctions as 
well as the spectra of elementary excitations are obtained from the solutions 
of the Gross-Pitaevskii equation including disorder.
Our work also shows that the presence of disorder has no significant impact on
the occurence of a bosonic phase transition for polaritons. This explains
why this phenomenon has been observed in a rather disordered system like CdTe.
This also gives good hope for the observation of such phase transition in 
even more disordered systems like GaN \cite{malAPL}. However, since disorder strongly 
affects the occurence of the superfluid phase transition, it could bring 
renewed interest in cleaner systems like GaAs based structures.

We thank K.V. Kavokin for enlightening discussions. We acknowledge the support of the 
STREP "STIMSCAT" 517769, and of the Chair of Excellence program of ANR.


\begin{thebibliography}{99}
\bibitem{ref1} A. Kavokin and G. Malpuech, Cavity Polaritons, Elsevier,
(2003).
\bibitem{Imamoglu} A. Imamoglu, J.R. Ram, Phys. Lett. A \textbf{214}, 193 (1996).
\bibitem{Dang98} L. S. Dang \emph{et al}. Phys. Rev. Lett \textbf{81}, 3920 (1998).
\bibitem{Yamamoto} H. Deng \emph{et al}. Science \textbf{298}, 199 (2002).
\bibitem{KAZ06} J. Kasprzak \emph{et al}. Nature \textbf{443}, 409 (2006).
\bibitem{KOS73} J. M. Kosterlitz and D. J. Thouless, J. Phys. C \textbf{6},
1181 (1973).
\bibitem{MAL03} G. Malpuech \emph{et al}. Semicond. Sci. \& Technol. \textbf{%
18}, Special issue on microcavities, edited by J.J. Baumberg and L. Vi\~na, S
395 (2003).
\bibitem{KEE06} J. Keeling, Phys. Rev. B \textbf{74} 155325 (2006).
\bibitem{FIS89} M. P. A. Fisher, P. B. Weichman, G. Grinstein and D. S.
Fisher, Phys. Rev. B \textbf{40}, 546 (1989).
\bibitem{Pitaevskiibook} L. Pitaevskii and S. Stringari, \textit{%
Bose-Einstein Condensation} (Oxford University Press, 2003).
\bibitem{EFR89} A. L. Efros and B. I. Shklovskii, \textit{Electronic
Properties of Doped Semiconductors} (Springer, Heidelberg, 1989).
\bibitem{explain1} Here and below we disregard the non-parabolicity effects on the cavity polariton dispersion which
are known to be small provided the temperature and polariton number is not too high~\cite{KEE06}.
\bibitem{explain2} We disregard the spin of polaritons and omit the spin degeneracy factor here and below.
\bibitem{Bagnato} V. Bagnato, D. Kleppner, Phys. Rev. A \textbf{44}, 7439-7441 (1991)
\bibitem{pitaevskii} E.M. Lifshitz and L.P. Pitaevskii, \textit{Statistical
physics}, part 2 (Pergamon Press, New York, 1980).
\bibitem{leggett} A.J. Leggett, Phys. Rev. Lett. \textbf{25}, 1543 (1970).
\bibitem{BER04} O.L. Berman, Y.E. Lozovik, D.W. Snoke, R.D Coalson, Phys.
Rev. B \textbf{70}, 235310 (2004).
\bibitem{KN} D.R. Nelson, J.M. Kosterlitz, Phys. Rev. Lett. \textbf{39},
1201 (1977). 
\bibitem{malAPL} G. Malpuech, A. Di Carlo, A. Kavokin, J.J. Baumberg, M. Zamfirescu and P. Lugli, Appl. Phys. Lett., 81, 412, (2002).
\end{thebibliography}
\end{document}